\documentstyle[prl,amsfonts,aps]{revtex}
\begin{document}
\title{\bf Multichannel Kondo screening in a \\ one-dimensional correlated
electron system}
\vspace{1.5em}
\author{Andrei A. Zvyagin$^{(a,b)}$, Henrik Johannesson$^{(a)}$ and Mats
Granath$^{(a)}$}
\address{$^{(a)}$ Institute of Theoretical Physics \\
Chalmers University of Technology and G\"oteborg University, S-412 96
G\"oteborg, Sweden} 
\address{$^{(b)}$ B. I. Verkin Institute for Low Temperature Physics and
Engineering \\ The National Academy of Sciences of Ukraine,
Kharkov 310164, Ukraine}
\maketitle

\begin{abstract}
We present the exact {\em Bethe Ansatz} solution of a multichannel model 
of one-dimensional correlated 
electrons coupled antiferromagnetically to a magnetic impurity of 
arbitrary spin $S$. 
The solution reveals that 
interactions in the bulk make the magnetic impurity
drive both spin {\em and} charge fluctuations, producing a mixed
valence $n_{imp} \neq 0$ at the impurity site, with an associated
effective spin $S_{eff} = S + |n_{imp}|/2$ in
the presence of a magnetic field. The screening of the impurity
is controlled by the size of the impurity spin independently of the number of
channels, in contrast to the multichannel Kondo effect for free electrons. 
\end{abstract}

{\em Introduction.}\ - \ Impurities play an essential role in correlated electron systems, 
in particular in 1D where even a small amount of defects may 
drastically change the properties of the system. While the effect of 
a single potential scatterer is by now fairly well understood 
\cite{pot}, the case of a {\em dynamical} scatterer in a 1D 
interacting electron system - like a magnetic impurity - largely 
remains an open problem \cite{LeeT}. 

Of particular interest is to understand what happens when the 
electrons are coupled to a local magnetic moment (of magnitude $S$) 
in several degenerate channels $m$. In the case of a 3D Fermi liquid, 
the impurity induces strong correlations among the 
electrons, with the low-temperature physics depending on the relation 
between the impurity spin and the number of electron channels 
\cite{NB,AndreiD,WieT,AL,sac}. It is natural to 
ask whether the correlations inherent in a 1D electron system ({\em 
Luttinger liquid,} \cite{Hal}) will modify 
this behavior. Does novel effects appear, or does one recover the same 
multichannel Kondo physics as for free electrons in 3D?  Apart from the possible 
experimental relevance of these
questions to the study of artificial impurities in mesoscopic devices
\cite{GK}, their resolution is an interesting issue in its own right. 
A study is also motivated by the recent interest in the Kondo effect in
the high-$T_c$ cuprates \cite{NagL}: it is well known that at least two
orbitals, $3d_{x^2 - y^2}$ and $3d_{z^2}$, play an essential role there,
so the Kondo effect should have multichannel nature. An analysis of the
simpler, analog problem in 1D - with {\em correlated} electrons - may provide
important clues for how to model the effect in the cuprates.

In this Letter we attack the problem by considering an {\em 
integrable} model of a magnetic impurity embedded into a 
multichannel interacting electron system. We here take a multichannel 
extension of the supersymmetric $t\!-\!J$ model in 1D 
\cite{tJ}, with the electrons coupled to a localized spin (of magnitude $S$) by an
antiferromagnetic exchange interaction that preserves 
integrability. This makes applicable a {\em Bethe Ansatz} 
approach, allowing us to obtain {\em exact} results for the 
groundstate properties as well as the finite temperature behavior. 
Our solution reveals that there are two distinct processes that 
govern the zero temperature response due to the impurity: First, 
the singlet Cooper-like pairs of electrons present in the ground 
state break up and get temporarily trapped when scattering off the 
impurity, thus producing an impurity valence $n_{imp} \neq 0$ with 
an associated effective impurity spin $S_{eff} = S + 
|n_{imp}|/2$ in the presence of a magnetic 
field $H$. Secondly, this effective composite spin gets screened by 
unbound electrons excited by $H$ from the spin singlet groundstate. 
The type of low-temperature behavior that emerges depends only on  
the size of the bare impurity spin $S$, and is insensitive to the number of channels.  
This property is different from that of the 
ordinary multichannel Kondo effect in a Fermi liquid.

{\em The model.}\ - \ The Hamiltonian ${\cal H}_{host}$ of the multichannel extension of the 
supersymmetric $t\!-\!J$ model can be written as ${\cal H}_{host} \equiv \sum_j
{\cal H}_{j,j+1}$, with  
\begin{equation}
{\cal H}_{j,j+1} = 
 - {\cal P}(c_{j,\sigma,f}^{\dagger}c_{j+1,\sigma,f} + h.c) 
{\cal P} - 
c_{j,\sigma,f}^{\dagger}c_{j,\sigma,f'}c_{j+1,\sigma',f'}^{\dagger}
c_{j+1,\sigma',f} + c_{j,\sigma,f}^{\dagger}c_{j,\sigma',f}
c_{j+1,\sigma',f'}^{\dagger}c_{j+1,\sigma,f'} \ .  
\label{Ham}
\end{equation}
Here $j=1,...,L$ is a site index, $\sigma = \pm$ denotes the spin 
projection, 
$f = 1,\dots,m$ is a {\em flavor} quantum number indexing the 
available electron channels, and ${\cal P}$ is a projector on 
the subspace of singly occupied states.
All indices are summed over. The scattering matrix for two electrons is given by 
${\hat X}(p_i-p_j) = [(p_i - p_j){\hat I} + i{\hat P}^{s}]\otimes 
[(p_i - p_j){\hat I} - i{\hat P}^{f}/[(p_i - p_j)^2 + 1]$ with 
$p_{i,j}$ the corresponding {\em charge rapidities}, and 
${\hat P}^{s}$ and ${\hat P}^{f}$ the permutation operators in 
spin and flavor subspace, respectively. The $\hat X$-matrices 
satisfy the Yang-Baxter equation ${\hat X}^{12}(p_1 - p_2)
{\hat X}^{13}(p_1 - p_3){\hat X}^{23}(p_2 - p_3) = {\hat X}^{23}
(p_2 - p_3){\hat X}^{13}(p_1 - p_3){\hat X}^{12}(p_1 - p_2)$, thus 
ensuring the integrability of the model. 

We now insert an additional site on the lattice, call it 0, attach a 
magnetic impurity to it, and couple it to the electrons on neighboring 
sites (sites 1 and L, given periodic boundary conditions). 
To preserve integrability this 
interaction must be judiciously chosen. We here follow the 
strategy pioneered in \cite{AndreiJ84} and {\em define} the 
impurity-host interaction via an electron-impurity S-matrix 
${\hat S}$ that satisfies the Yang-Baxter equation ${\hat X}^{12}
(p_1 - p_2){\hat S}^{10}(p_1 - p_0){\hat S}^{20}(p_2 - p_0) = 
{\hat S}^{20}(p_2 - p_0) {\hat S}^{10}(p_1 - p_0){\hat X}^{12}(p_1 
- p_2)$, with $p_0$ measuring the impurity energy level. This 
approach assures that the model remains integrable in presence of 
the impurity. The ${\hat S}$-matrix can still be chosen in a 
number of ways \cite{mag}, and here we take it to be similar to 
that of the multichannel Kondo problem in a free electron gas 
\cite{sac} and let it act nontrivially only in the spin subspace. 
Writing out the components,
\begin{equation}
{\hat S}^{\alpha \alpha'}_{M M'}(p) = a(p){{(p + i/2)
\delta_{\alpha \alpha'}\delta_{MM'} + i{\sigma}^k_{\alpha 
\alpha'}S^k_{MM'} }\over {p+i}} \ , \label{scatt} 
\end{equation}
with ${\sigma}^k$ the Pauli matrices $(k\!=\!x,y,z)$, $S^k$ the 
impurity spin matrices, $|M| \le S$ the component of the impurity 
spin $S$ (with unprimed/primed indices referring to 
incoming/outgoing states), and $a(p) \equiv [(p^2 + 1)/(p^2 + 
(S + 1/2)^2)]^{1/2}$. The corresponding impurity 
Hamiltonian ${\cal H}_{imp}$ can be written in the form: 
\begin{eqnarray}
{\cal H}_{imp} &  = & J\bigl({\cal H}_{L,0} + {\cal H}_{0,1} + 
\{ {\cal H}_{L,0},{\cal H}_{0,1} \}\bigr) + (1 - 3S(S+1)J - 
{J\over 4}){\cal H}_{L,1}  \nonumber \\ 
& + & 2p_0J[({\cal H}_{L,0} + {\cal H}_{0,1}),{\cal H}_{L,1}] \ . 
\label{Himp}
\end{eqnarray}
Here ${\cal H}_{0,1}$ and ${\cal H}_{L,0}$ are generalized permutation operators of a
particle with spin $S$ but no flavor (impurity) and an electron (carrying both spin {\em
and} flavor), while ${\cal H}_{L,1}$ is defined in Eq. (1).
The function $J = [p_0^2 + (S + {1\over2})^2]^{-1}$ plays the 
role of an effective exchange constant between the impurity site 
and neighboring sites. 
We should mention that the structure
of ${\cal H}_{imp}$ for the periodic chain in (\ref{Himp}) simplifies
considerably when the impurity is located at the edge of an {\em open} 
chain, with the impurity connected to the host via only one link, with 
coupling constant $J$. It is here worth pointing out that the low-energy behavior 
of impurities in integrable open and periodic chains of correlated electrons 
are qualitatively the same,
as was shown recently in \cite{Z}.
We should also point out that our approach is different from that recently advocated 
by Wang and Voit in their study of the ferromagnetic Kondo effect 
in a Luttinger liquid, where the impurity is simulated by a 
boundary potential \cite{WV}. In contrast, in our formulation 
the full dynamics of the impurity is retained. 

{\em Bethe Ansatz equations.}\ - \ The model thus constructed can be diagonalized exactly by an 
{\em algebraic Bethe ansatz} \cite{qism}. The procedure is rather cumbersome, 
and we here only give the result. The eigenstates are characterized 
by $m+2$ sets of quantum numbers: The {\em charge rapidities} 
$\{p_j\}_{j=1}^N$ (with $N$ the total number of electrons), the 
{\em spin rapidities} $\{\lambda_{\alpha}\}_{\alpha =1}^M$ ($M$ 
counting the number of spin-down electrons), and the $m$ sets of 
{\em flavor rapidities} $\{\mu_{\beta}\}_{\beta=1}^{m^{(i)}}, 
i=1,2,...,m$ (with $m^{(i)} = \sum_{k=i+1}^{m} n^{(k)}, \ n^{(k)}$ 
counting the number of electrons in channel $k, k=1,2,...,m$). 
Each state corresponds to a particular solution of the nested 
{\em Bethe Ansatz} equations 
\begin{eqnarray}
&&\prod_{\tau =\pm}\prod_{\beta=1}^{m^{(k+\tau)}}e_1
(\mu_{\alpha}^{(k)} - \mu_{\beta}^{(k+\tau)}) = 
\prod_{\gamma=1}^{m^{(k)}}e_2(\mu_{\alpha}^{(k)} - 
\mu_{\gamma}^{(k)}) \ , \nonumber \\
&&[e_{2S+1}(p_j - p_0)e_2(p_0 - p_j)]^{1/2}e_{1}^{L}(p_j) = 
\prod_{\alpha =1}^{M}e_1(p_j - \lambda_{\alpha})\prod_{\beta 
=1}^{m^{(1)}}e_1(\mu_{\beta}^{(1)} - p_j) \nonumber \\ 
&&e_{2S}(\lambda_{\alpha} - p_0)\prod_{j=1}^{N}e_1(\lambda_{\alpha} 
- p_j) = \prod_{\beta=1}^{M}e_2(\lambda_{\alpha} - \lambda_{\beta}) 
\ ,  \label{BAE} 
\end{eqnarray}
where $e_n(x) \equiv (2x +in)/(2x -in)$,  $\mu_{j}^{(0)} = p_j$, 
$m^{(0)} = N$, $m^{(m+1)} =0$, and $L$ is the number of lattice 
sites (not counting the impurity site). 
We shall assume that no external fields couple to the electron 
channels, and can hence confine our attention to the flavor-singlet 
subspace. The energy (up to an additive constant) and the 
magnetic moment are then equal to $E = \sum_{j}^{N}(p_{j}^2 + 
(1/4))^{-1}$ and $S^z = N/2 + S - M$, respectively. 

The thermodynamics of the model is described by the usual string 
hypothesis  \cite{YY}. In the thermodynamic limit, with $L, N, M, 
m^{(j)} \to \infty$, their ratios being fixed, the model has the 
following possible excitations: (i) unbound electrons with charge 
rapidities $p_j$; (ii) spin singlet Cooper-like pairs with  $p_j 
= \lambda_j \pm i/2$; (iii) spin strings 
(bound states of any number of ``down spins'');  
and (iv) flavor strings for each channel.
(Since the host interactions in (\ref{Ham}) 
carry opposite signs in spin- and flavor subspaces, bound states 
of different channels as well as bound states of electrons and 
spin strings are suppressed.) Introducing distribution functions 
{\em (densities)} for particles and holes for each class of 
excitations we can write down the corresponding thermodynamic 
{\em Bethe ansatz} equations. Then, by minimizing the free energy, 
we extract the integral equations for each excitation class. 
These equations have the same structure as those of the 
multichannel Kondo problem for free electrons \cite{Tsv}, and 
differ only in the driving terms (which are independent of energy 
and density). This implies that in the high-temperature limit, 
where the driving terms are unimportant, the effect of our 
impurity is similar to that of a Kondo impurity in a 
multichannel free electron gas. 

{\em Groundstate properties.}\ - \ Let us focus instead on the groundstate properties of the 
impurity, and study how it depends on the host band filling and 
an applied magnetic field. In the zero temperature limit, $T \to 
0$, only excitations of classes (i), (ii), and (iv) (for unbound 
{\em flavorons} and pairs of them, i.e. flavor strings of length 
1 and 2 for each channel) can have negative energies. Thus, the 
filling of these states is determined by the Dirac seas of the 
groundstate of the model. In the singlet flavor sector we can 
solve the equations for densities and energies of the flavorons 
as functions of the densities and energies of unbound electrons 
and singlet spin-charge pairs. As a result, we obtain the ground 
state equations for unbound electrons and pairs:
\begin{eqnarray}
\rho_h(p)\! +\! (1\! -\! a_1\! \star \!s_1)\! \star\! (\rho (p)\! +\! a_1 \!\star\! 
\sigma (\lambda))&\! =\!& a_1 (p)\! +\!
{\textstyle \frac {1}{2L}}[a_{2S+1}\! -\! a_2](p\! -\! p_0) \ , \nonumber
\\
\sigma_{h}(\lambda) + (1 + a_2)\star (1 - a_1 \star s_1) \star 
(\sigma (\lambda)\! +\! s\! \star \!\rho (p))&\! =\!& a_{2}(\lambda)\! + \!
{\textstyle \frac {1}{2L}} [a_{2S+2}\! - \!a_{2S}\! -\! a_3\! -\! a_1]
(\lambda\! -\! p_0) \ , 
\label{groundBA}
\end{eqnarray}
where $\star$ denotes convolution, $\rho (\rho_h)$ and $\sigma 
(\sigma_h)$ are densities for unbound electrons and pairs, 
respectively. The integration intervals are given by $|p| > B$ 
and $|\lambda | > Q$, with  $Q$ and $B$ playing the role of 
Fermi points for unbound electrons and pairs, respectively. The 
Fourier transforms of the kernels $a_n$, $s$ and $s_1$ are given 
by $\exp (-n |\omega |/2)$, $\cosh^{-1}/2 \omega$, and $\sinh 
[(m-1)|\omega |/2] / \sinh (m|\omega |/2)$, respectively. 
Eqs. (\ref{groundBA}) are linear in the densities, and the 
driving terms of the host and the impurity are additive. 
Separating the densities into bulk and impurity parts then allows 
us to calculate the valence and the magnetization of the impurity 
in the groundstate. In the absence of an external magnetic field 
we have $B = \infty$ (no unbound electrons). The limit $Q \to 
\infty$ corresponds to a vanishing pair density, while $Q \to 0$ 
is the limit of $1/2m$-filled bands of the 
host (no pair holes, corresponding to a vanishing  
Fermi velocity). For the nonmagnetic groundstate it 
follows that the effective valence of the impurity varies as a 
function of electron number from $n_{imp} = 0$ for vanishing pair 
density to $n_{imp} = - m$ for $1/2m$-filled 
bands, with the valence measured w.r.t. the groundstate of 
spin-paired electrons (i.e. its negative sign indicates an excess 
of pair {\em holes} due to the presence of the impurity). This 
unusual behavior is caused by the host interaction in (\ref{Ham}) which supports a 
groundstate with a Dirac sea of bound singlet Cooper-like 
electron pairs, not present in the free electron gas. 

Of special interest is the behavior of the impurity 
magnetization $S^z_{imp}$. With no magnetization in the bulk, 
we have $S^z_{imp} = S$. By turning on a magnetic field, 
the number of unbound electrons increases while the 
number of singlet pairs decreases (as required by electron number 
conservation). By eliminating the pair density $\sigma (\lambda)$ 
from the second equation in (\ref{groundBA}), we obtain the 
Fredholm equation 
\begin{equation}
\rho_h (p) \!+\! \rho (p) \!-\!F \star \rho = s \star \sigma_h \!+\! 
s(p) \!+\! {\textstyle \frac {1}{2L}} s \star a_{2S}(p - p_0) \ , 
\label{rho}
\end{equation}
where $F(\omega) = 1 - \tanh (|\omega|/2)[1 - \exp(-m|\omega|)]^{-
1}$, 
and the integration over the pair hole density is over the finite 
interval $[-Q, Q]$. This yields an explicit connection between the 
densities of unbound particles (electrons or holes) and spin singlet 
pair holes. As the Zeeman splitting is typically much smaller than 
the Fermi energy, we can neglect the influence of the pairs on the 
impurity magnetization as long as the magnetic field is 
sufficiently weak. For this case $(H \ll 1)$, choosing $Q = 0$ (i.e. 
$1/2m$-filled bands) and assuming that $|p_0| 
\gg B$, we can solve Eq. (\ref{rho}) exactly by reducing it to a 
sequence of coupled Wiener-Hopf integral equations. In this way we 
obtain {\em two} distinct regimes for the behavior of $S^z_{imp}$ with 
magnetic field. If $S > 1/2$ 
the impurity spin becomes asymptotically free:   
\begin{equation}
S_{imp}^z =\mu \bigl[1 \pm {\textstyle \frac {m}{2}}(|\log(H/T_H)|)^{-
1} 
+ \dots \bigr] \ , \ \ S > \frac{1}{2} \label{magn} 
\end{equation}
with $T_H = 2\pi ({\textstyle \frac {m}{2e}})^{m/2}T_K / \Gamma 
({\textstyle \frac {m}{2}})$, and $T_K \propto \exp (-\pi p_0)$ 
playing the role of the Kondo temperature. When $H \ll T_H$, $\mu = 
S$, 
and the upper sign in (\ref{magn}) gets selected. On the other hand, 
with $T_H \ll H \ll 1$,  $\mu = S + m/2$ and the 
lower sign in (\ref{magn}) is selected. Note that for this case $H$ 
is still much smaller than the spin saturation field in the bulk 
(corresponding to a transition to a ferromagnetic bulk state). Also note
that the {\em underscreened} behavior in (\ref{magn}) for $H << T_H$ 
appears only for $m<2S$ in a free electron gas \cite{sac}, in contrast to the
correlated host studied here. For the 
{\em exactly screened case} $S = 1/2$, 
\begin{equation}
S_{imp}^z \sim \frac{mH}{2\pi T_K} , \ \ S=\frac{1}{2},
\end{equation}
producing a finite zero-field 
magnetic susceptibility. Again, this is different from a free electron 
host where this behavior sets in for $m=2S$. Most importantly, the case of 
{\em overscreening} ($m > 2S$ for a free host, with critical scaling 
of the impurity magnetization) {\em does not exist} in this 
interacting electron system. Thus, electron correlations in
the host due to direct electron-electron interaction here suppresses
the critical overscreened behavior of a Kondo-like impurity, similar to what
happens in the presence of a channel anisotropy for free electrons \cite{AndJ}.
It is important to realize that by increasing the magnetic 
field we deplete the number of singlet pairs. Therefore, as we have 
already seen, by varying the field we can smoothly tune the 
effective impurity valence $n_{imp}$ from $- m$ to 0. In the 
underscreened case this means that the uncompensated part of the 
{\em effective} impurity spin $S_{eff} = S +  
|n_{imp}|/2$ decreases with increasing field, and recovers its 
zero-field value $S$ at the spin-saturation field. 

{\em Low-temperature thermodynamics.}\ - \ We now turn to the low-temperature thermodynamics. 
As we have 
already mentioned, the thermodynamic Bethe ansatz equations 
differ from those for the free electron gas multichannel Kondo 
problem only in the driving terms, and we can hence use an 
analysis similar to that in \cite{AndreiD,Tsv}. The main idea is 
to rewrite the $T \to 0$ thermodynamic Bethe ansatz equations for 
{\em all} excitations on universal form. This can be 
realized in the ``Kondo limit'' (with suppressed charge fluctuations), 
yielding for the universal energy potentials $\phi_j$ 
\cite{Tsv}:
\begin{equation}
\phi_j (\lambda ) = s \star \log (1 + e^{\phi_{j-1}})(1 + 
e^{\phi_{j+1}}) - \delta_{j,m}e^{\pi \lambda} \ , \label{therm}
\end{equation} 
with $\lim_{j \to \infty}(\phi_j /j) = H/T$. It follows that in 
this limit the impurity low-temperature free energy can be written as a
function of $\phi_{m}$ : 
\begin{equation}
F_{imp} = - T \int_{-\infty}^{\infty} {{d\lambda a_{2S}(\lambda)
\log (1 + e^{\phi_{m}})}\over 2 \cosh [\pi\lambda + \log (T/T_K)]} 
\ .
\end{equation}
The qualitative behavior of $F_{imp}$ is independent of the relative values
of $m$ and $S$, in analogy with the groundstate properties.  
In the underscreened case $S>1/2$, the impurity magnetic susceptibility $\chi_{imp}$ 
shows Curie-like temperature dependence while the specific heat $C_{imp}$ 
exhibits a Schottky anomaly at $T \sim H$ and a Kondo resonance. For 
the exactly screened case $S = 1/2$ local 
Fermi liquid behavior holds: ${\chi}_{imp} \approx m/2\pi T_K$, $C_{imp} 
\approx \pi m T/(m+2)T_K$. 

{\em Discussion.}\ - \ To conclude, we have found an exact {\em Bethe Ansatz} solution of 
the problem of a Kondo-like antiferromagnetic impurity embedded 
into the multichannel supersymmetric $t\!-\!J$ model.  
Our solution reveals that there are two mechanisms governing the 
magnetic behavior of the impurity: holes in the distribution of 
Cooper-like pairs increase the effective spin of the impurity, 
while unbound electrons screen this effective value. Note, that 
for the $t\!-\!J$ model there is no spin gap, so the unbound 
electrons and Cooper-like pairs coexist even for small external 
magnetic fields. In contrast to the multichannel Kondo effect 
in a Fermi liquid, the screening with unbound electrons are 
realized in only {\em two} different ways independent of the number 
of channels, and controlled only by the value of the {\em bare} impurity 
spin: {\em underscreening}, with an asymptotically free effective 
impurity spin, and {\em exact screening} with a finite magnetic 
susceptibility of the impurity. Overscreened critical 
behavior is absent for this correlated electron model.
As required by integrability, our model supports only
forward scattering of the electrons off the impurity,
and hence does not produce localization of
the electrons. In contrast, more realistic 1D magnetic impurity
models must contain also back scattering terms which are
expected to renormalize to an effective boundary potential,
in addition to producing an impurity thermal response sensitive
to the electron-electron interaction \cite{GJ}.
Yet, our 
model dramatically shows how correlations in the bulk make 
the magnetic impurity drive both spin {\em and} charge 
fluctuations. This property is expected to be generic and our 
exact solution provides a detailed picture of its possible 
realization.  

A. A. Z. gratefully thanks the Institute of Theoretical Physics 
at Chalmers and G\"oteborg University for kind hospitality. H. J. acknowledges 
support from the Swedish Natural Science Research Council.

\end{document}